\documentclass[a4paper,fleqn,usenatbib, useAMS, onecolumn]{mnras}

\usepackage[T1]{fontenc}
\usepackage{ae,aecompl}

\usepackage{graphicx}	
\usepackage{amsmath}	
\usepackage{amssymb}	

\title[Transfer equations for polarized radiation]{Cyclotron line formation in the magnetized atmospheres of compact stars: I. The transfer equations for polarized radiation}
\author[M. A. Garasev, E. V. Derishev, Vl.V. Kocharovsky and V. V. Kocharovsky]{M. A. Garasev$^{1, 2}$\thanks{E-mail:
garasyov@gmail.com}, E. V. Derishev$^1$, Vl.V. Kocharovsky$^1$ and V. V. Kocharovsky$^{1, 3}$\\
$^{1}$Institute of Applied Physics, 46 Ulyanova Str., 603950 Nizhny Novgorod, Russia \\
$^{2}$Lobachevsky State University of Niznhi Novgorod, 23 Gagarin Avenue, 603950 Nizhny Novgorod, Russia \\
$^{3}$  Department  of Physics and Astronomy, Texas A\&M University, College Station, TX 77843-4242, USA
}

\date{Accepted XXX. Received \today; in original form \today}

\pubyear{2015}

\begin{document}
\label{firstpage}
\pagerange{\pageref{firstpage}--\pageref{lastpage}}
\maketitle

\begin{abstract} 
{
We find the forms of the transfer equations for polarized cyclotron radiation in the atmospheres of compact stars, which are simple enough to allow practical implementation and still preserve all important physical effects. We take into account a frequency redistribution of radiation within the cyclotron line as well as the relativistic and quantum-electrodynamic effects. Our analysis is valid for the magnetic fields up to $10^{13}$\,G and for temperatures well below 500\,keV.} We present and compare two forms of the radiation transfer equations. The first form, for the intensities of ordinary and extraordinary modes, is applicable for the compact stars with a moderate magnetic field strength up to $10^{11}$\,G for typical neutron star and up to $10^9$\,G for magnetic white dwarfs. The second form, for the Stokes parameters, is more complex, but applicable even if a linear mode coupling takes place somewhere in the scattering-dominated atmosphere. Analysing dispersion properties of a magnetized plasma {in the latter case, we describe a range of parameters where the linear mode coupling is possible and essential.
 }
\end{abstract}

\begin{keywords}
 line: formation -- radiative transfer -- scattering -- stars: neutron -- white dwarfs.
\end{keywords}



\section{Introduction}

An emission from the atmospheres of compact stars (neutron stars and white dwarfs) can provide a valuable information on the physical properties of plasma surrounding these objects. {The most universal feauture of that emission is the cyclotron line, whose formation was studied by many authors (e.g., \citet{m45, m4, m_n1, m22_z}).} A significant progress in the X-ray observations, facilitated by the space-based observatories like {\it XMM-Newton} and {\it Chandra}, revives an interest in the theoretical and observational studies of a line formation in the atmospheres of such stars  \citep{m1, m2, m3, m3_1}. A discovery of the spectral features in the thermal spectra of the X-ray dim isolated neutron stars and in the central compact objects (CCOs) is especially interesting. {One of the possibilities is that they are the cyclotron lines forming in the outer layers of the atmospheres (see, e.g., \citet{m35, m35_a, m35_b, m35_c})}. 

The physical parameters in the upper (near-photosphere) layers of the compact stars' atmospheres are such that scattering of photons strongly dominates over their absorption \citep{m4, m6}, that is an electron collision rate is much smaller than a spontaneous radiative rate. In this case, a population of the Landau levels is governed by a local number density of the resonant photons. Usually, an occupation number of resonant photons in the upper atmosphere is much smaller than a thermodynamic equilibrium value, so that the excited Landau levels are  underpopulated to such an extent that almost all electrons reside at the ground level and an effective transverse temperature is nearly zero \citep{m27}. The longitudinal momentum distribution of electrons is supposed to be Maxwellian (in line with most of authors), though in the upper layers of the atmosphere, where the plasma is almost collisionless, this assumption has to be taken with care. 

A strong magnetic field of the neutron stars makes the quantum-electrodynamic (QED) effects also important \citep{m7}. A number of authors investigated in detail an influence of  the QED effects on the dispersive characteristics of electromagnetic waves in a magnetized plasma (see, e.g., \citet{m7, m8, m9}). It was shown that a vacuum polarization can affect the radiation transfer by modifying a polarization of the normal waves (modes) and inducing the additional features in their opacity coefficients \citep{m10, m32, m42}. Also, it was indicated in \citet{m11}, that, due to a linear mode coupling, a description of a radiative transport within the framework of geometrical optics (which is widely used in lots of calculations of a radiation transfer in the atmospheres of compact stars) is, in general, inadequate near the points, where the vacuum contribution to the refraction indices becomes comparable with the plasma one, i.e., in the vicinity of the so-called 'vacuum resonance' points. Below we show that, even in a relatively weak (for  the neutron stars) magnetic  field of the order of $B\sim10^{11}G$, the vacuum polarization can significantly modify a cyclotron line profile in the emergent spectra.

A major part of  a prior work on the modeling of cyclotron lines in the spectra of highly magnetized compact stars has been done either within a cold plasma approximation \citep{m12, m13, m14} or  within a limit of a nonrelativistic cyclotron resonance condition 
\citep{m15, m16, m17}. The latter approach greatly simplifies a solution since  the radiation transfer equations can be reduced to the quasi-monochromatic ones by a proper choice of variables. As it was firstly mentioned in \citet{m5}, at each resonant scattering there is a conserved quantity --  a velocity of an electron, which is in resonance with a given photon,
\begin{equation}
\beta = \frac{\omega-\omega_B}{\omega\cos\theta},
\label{eq1}
\end{equation}
where $\omega$ is the frequency of a photon, $\theta$ the angle between the photon wave vector and the magnetic field, $\omega_B = eB/m_ec$ the cyclotron frequency, and $\beta$ the velocity of electron in the units of the speed of light $c$. This allows one to simplify a radiation transfer problem by using a velocity $\beta$ instead of a frequency $\omega$ as an independent variable. Such an approach proved to be very helpful in the limit of low optical depths and was widely used for studying the cyclotron line formation in the neutron star magnetospheres \citep{m18, m19}. In what follows, we refer to that approximation as a quasi-coherent approximation, because the changes in frequencies due to a Doppler effect do not accumulate and all resonant photons remain locked in the line core.

Recently we showed \citep{m21, m22} that, for the extremely magnetized white dwarfs and for  the neutron stars with magnetic fields stronger than $10^{10}$\,G, the relativistic corrections to the Doppler resonant conditions become important and brake the quasi-coherent approximation, leading to an efficient escape of the cyclotron radiation from a relatively large optical depth. 
{In their most general form, the transfer equations for a magnetized plasma + vacuum medium (see, e.g., \citet{m22_x}) are too complicated even for numerical simulations. }

{
The goal of this paper is to find the simplest possible forms of transfer equations, which would be simple enough for practical implementation and, at the same time, still include all the relevant physical effects and provide accurate results. The minimal set of physical effects to be incorporated into the transfer equations depends on the physical conditions in the atmosphere of compact stars. We explore the parameter space for a fully ionized magnetized plasma in a mildly-relativistic approximation (with temperature in energy units $T \ll m_ec^2$) with moderate (up to $10^{13}$\,G) magnetic fields, and identify which form of the transfer equations is suitable for which parameters. Our formulation of the radiation transfer problem allows for the effects of the vacuum polarization as well as for a non-trivial redistribution of photons over frequencies and their escape from the cyclotron line due to the relativistic Doppler effect in the scattering-dominated atmospheres. }

{We concentrate on describing correctly the scattering of photons near the fundamental harmonic of cyclotron resonance.  This harmonic is the most prominent feature in the spectrum, and radiation pressure at the fundamental frequency is a major factor \citep{m22_s, m22_t, m22_w}, which determines the structure of the atmospheres and possible formation of radiation-driven winds. For this reason, we do not consider higher harmonics, but adding them is straightforward. }

The paper is organized as follows. In the next section, we derive all the coefficients in a complete set of the radiation transfer equations for the incoherent mode intensities. The coefficients are expressed through the dielectric tensor for the 'warm mildly-relativistic plasma + magnetized vacuum' media, which is also calculated. That set of equations can be used only in the case of the geometrical optics, where the radiation transfer in an inhomogeneous plasma is adiabatic, i.e., preserves a wave's mode composition in the absence of scattering. In the third section, we derive a more general but more complex set of the transfer equations for the Stokes parameters. They are to be used when the adiabatic approximation is not valid. In the last section, we discuss the range of plasma parameters, where the radiation transfer can be non-adiabatic, i.e., can undergo the linear mode coupling. We show that the vacuum resonance zone for the cyclotron radiation is located at the higher optical depths as compared to the continuum and may lie far below the photosphere. We also indicate the three domains in the parameter space, related to the three approximations useful in the radiation transfer problem. In the discussion section, we outline the most important results of the work and their possible astrophysical implications.

\section{Transfer equations for the X-, O-mode intensities}

The simplest, yet widely employed way to describe the transfer of a polarized radiation in the magnetized plasma is to use the transfer equations for the intensities of two electromagnetic modes -- ordinary(O) and extraordinary (X) ones. In a tenuous medium, where a difference of the dielectric tensor from its vacuum value is small, 
any electromagnetic wave is almost transverse and can be easily represented as a superposition of two modes. The magnetoactive plasma is a tenuous medium at frequencies well above the plasma frequency $\omega_\mathrm{L} = \sqrt{4\pi e^2N_e/m_e}$ (and outside of strong resonances), where $e$ is the elementary charge, $N_e$ the number density of electrons, and $m_e$ the mass of an electron. In this case the scattering phenomena became transparent if described in the form of two incoherent modes. The extraordinary mode has a resonance in opacity in the vicinity of cyclotron harmonics. The ordinary mode also has  a resonance, but it is much weaker and completely disappears in the limit of a cold plasma. In the geometrical optics (adiabatic) approximation, these modes are coupled with each other only through scattering. 

The radiation transfer equations in this case are \citep{m36}
\begin{equation}
\frac{\mathrm{d}J_i}{\mathrm{d}s} = -(\sigma^\mathrm{(sc)}_i+\sigma_i^\mathrm{(abs)})J_i + \int\sum_{j=1}^{2}\frac{\omega}{\omega'}\frac{\mathrm{d}\sigma^\mathrm{(sc)}_{ji}}{\mathrm{d}\Omega \mathrm{d}\omega}(\omega', \Omega' \rightarrow \omega, \Omega)J_j(\omega', \Omega') \mathrm{d}\Omega'\mathrm{d}\omega' + \frac{1}{2}B_{\omega, T}\sigma^\mathrm{(abs)}_i.
\end{equation}
Here $J_i(\Omega, \omega, s)$ is the intensity of radiation in the $i$-th mode with a unit polarization vector $\mathbf{e}_i$ related to a unit spectral interval and a unit solid angle (a usual convention is $i=1$ for the extraordinary mode and $i = 2$ for the ordinary mode), $\mathrm{d}\Omega$ the elementary solid angle, $s$ the coordinate along the ray, $\mathrm{d}\sigma^\mathrm{(sc)}_{ji}/\mathrm{d}\Omega \mathrm{d}\omega$ the differential scattering cross section, which describes scattering from mode $j$ to mode $i$, $\sigma_i^\mathrm{(abs)}$ the coefficient of the free-free absorption for the $i$-th mode, $ \sigma^\mathrm{(sc)}_i = \int{\sum_{j=1,2}\mathrm{d}\sigma_{ij}/(\mathrm{d}\Omega \mathrm{d}\omega) \mathrm{d}\Omega \mathrm{d}\omega} $ the total scattering cross-section, and
\begin{equation}
B_{\omega, T} = \frac{\hbar\omega^3}{4\pi^3c^2}\frac{1}{\exp(\hbar\omega/T)-1}
\end{equation}
the thermal equilibrium (Planck) intensity. The primed quantities are for the incident wave, those without prime are for the scattered wave. {Here and below $T$ is a temperature measured in energy units. }

 A photon redistribution over frequencies in the cyclotron line is, in general, an important effect \citep{m22}, which can lead to a significant (by several orders of magnitude) increase in the escape probability for the resonant radiation from large optical depths. The frequency redistribution in the cyclotron line is due to the relativistic Doppler effect. A motion of electrons in the strong magnetic field is essentially unidirectional, and, hence, the electrons resonantly interact only with the photons of almost certain frequency. The resonance condition for scattering by an electron, which undergoes a transition between the ground level and the $ \ell $-th Landau level, takes the form \citep{m9}
\begin{equation}
\frac{\hbar\omega}{m_ec^2}+\sqrt{1+P^2}-\sqrt{1+(P+\hbar k_{||})^2+2b\ell} = 0,
\label{eq2}
\end{equation} 
where $P$ is the momentum of an electron along the magnetic field (in the units of $m_ec$),  $k_{||}$  the component of the photon's wave vector along the magnetic field direction (in the units of $m_ec/\hbar$), $b = B/B_\mathrm{cr}$  the magnetic field strength (in the units of Schwinger field $B_\mathrm{cr} = m_e^2c^3/(\hbar e)$). Eq.(\ref{eq2}) has two roots, whereas the non-relativistic resonance condition (\ref{eq1}) has only one. The use of only one root, instead of two, does not allow for the actual redistribution of photons over frequencies, i.e., for a change in a resonant group of electrons, and prevents from their escape from the cyclotron line core. This leads to inaccurate results even if the plasma temperature is far below the relativistic values. 

The simplest option to preserve the two roots in the resonance condition is to expand Eq. (\ref{eq2}) into a power-law series, keeping terms up to the second order in $\beta$:
\begin{equation}
\omega(1-\beta\cos\theta-\frac{\omega}{2\omega_B}b\cos^2\theta) = \ell\omega_B(1-\frac{\beta^2}{2}-\frac{1}{2}b\ell).
\label{eq3}
\end{equation}
The resonance condition in such a form describes all essential effects near the cyclotron resonance {in a mildly-relativistic plasma. In what follows in this section we present the scattering properties and the wave characteristics in a warm plasma. }

Let us assume that the electron distribution function over the longitudinal velocities is Maxwellian,
\begin{equation}
f(\beta) =\frac{1}{\sqrt{2\pi}\beta_T}\exp{\left\{-\frac{\beta^2}{2\beta_T^2}\right\}},
\label{eqx4}
\end{equation}
where $\beta_T = \sqrt{T/(m_ec^2)}$ is a typical thermal velocity of electrons (in the units of the speed of light).
Also we assume that all the electrons are at the ground Landau level and consider only excitations to the first Landau level, that is $\ell = 1$ in Eq.~(\ref{eq3}). Usually it is safe to neglect the population of excited Landau levels in the upper layers of compact stars' atmospheres \citep{m27}. The reason is that the rate of spontaneous emission is many orders of magnitude higher than the rate of collisional excitation. 

To derive the transfer equations, which take into account the frequency redistribution of photons, let us start from the general expressions for the differential scattering cross-section \citep{m37}
\begin{equation}
  \frac{\mathrm{d}\sigma^\mathrm{(sc)}_{ji}}{{\mathrm{d}\Omega \mathrm{d}\omega}}(\omega', \Omega' \rightarrow \omega, \Omega) = r_0^2\frac{\omega'}{\omega}\int\mathrm{d}\beta f(\beta)\left|\left\langle\mathbf{e'}_j|\hat{\Pi}|\mathbf{e}_i\right\rangle\right|^2\delta(\omega-\omega' + \Delta\omega), 
\end{equation}
where $r_0 = e^2/(m_ec^2)$ and 
\begin{equation} 
\Delta\omega =-\omega\beta\cos\theta + \omega'\beta\cos\theta'-b\left(\frac{\omega}{\omega_B}\cos\theta-\frac{\omega'}{\omega_B}\cos\theta'\right)^2.
\end{equation}
It is also convenient to express the total scattering cross-section in the same notation,
\begin{equation}
\sigma^\mathrm{(sc)}_i =  \frac{8\pi}{3}r_0^2\int f(\beta)\mathrm{d}\beta\left|\hat{\Pi}\mathbf{e}_i\right|^2,
\end{equation} 
and through the optical theorem
\begin{equation}
\sigma^\mathrm{(sc)}_i = -4\pi r_0\frac{c}{\omega}\mathrm{Im}\langle \mathbf{e}_i | \hat{\Pi} | \mathbf{e}_i \rangle.
\label{eq_sigma}
\end{equation}

 In the dipole approximation, the matrix elements of the scattering amplitude for an individual electron is equal to
\begin{align}
   \Pi_{xx}&=\Pi_{yy} = -\frac{\omega}{2}\left(\frac{1}{\omega + \omega_B} + \frac{1}{\omega^*-\omega_B^*}\right), \\
	 \Pi_{xy}&=-\Pi_{yx} = -\frac{\mathrm{i}\omega}{2}\left(\frac{1}{\omega + \omega_B} - \frac{1}{\omega^*-\omega_B^*}\right), \\
	 \Pi_{xz}&=\Pi_{zx} = -\mathrm{i}\Pi_{yz} = \mathrm{i}\Pi_{zy} = -\frac{\omega}{2\omega_B}\left(\frac{\omega\beta\sin\theta}{\omega^*-\omega_B^*}\right), \\
	 \Pi_{zz}&=\frac{\omega}{\omega + \mathrm{i}\gamma}; \\
	\omega^* &= \omega(1-\beta\cos\theta-\frac{\hbar\omega}{2m_ec^2}\cos^2\theta) + \mathrm{i}\gamma, \\
	\omega_B^* &=\omega_B(1-\beta^2/2-\frac{b}{2}).
\end{align}
These equations are derived as a mildly-relativistic limit from the general form which can be found in \citet{m29} (equation 5.3.5). We use the reference frame with z-axis along the magnetic field and x-axis in the plane, containing the magnetic field and the wave vector (the same reference frame will be used below unless specified otherwise). In the above expressions,
$\gamma = \gamma_\mathrm{r} + \gamma_\mathrm{ff}$ is the total resonance width for an individual particle, which takes into account both the radiative, $\gamma_\mathrm{r}$, and collisional, $\gamma_\mathrm{ff}$, broadening\footnote{A more refined, but still approximate, methodology calls for the use of two different collisional frequencies for the longitudinal and transverse motions \citep{m40}. Since this difference can hardly lead to an observable effect in the scattering-dominated atmospheres, we adhere to a simpler approach.}. Based on the fact that these partial line widths are related to the total scattering cross section and to the free-free absorption, respectively, \citet{m39} deduced that the ratio between the scattering cross section and the absorption cross section equals to the ratio of the corresponding line widths. Thus, the cross section for the free-free absorption is
\begin{equation}
  \sigma^\mathrm{(abs)}_i = \frac{\gamma_\mathrm{ff}}{\gamma_\mathrm{r}}\sigma^\mathrm{(sc)}_i.
\end{equation} 
In the non-relativistic limit, $\gamma_\mathrm{r} = (2/3)e^2\omega^2/(m_ec^3)$, where $\alpha_\mathrm{f} = e^2/(\hbar c)$ is the fine structure constant. {In the quantum approach, which is necessary to use in the stronger magnetic fields, the radiative damping rates was calculated in \citet{dop1}. }A detailed evaluation of $\gamma_\mathrm{ff}$ can be found in \citep{m40}; in the most interesting frequency range near the cyclotron frequency,
\begin{equation}
  \gamma_\mathrm{ff} \approx \sqrt{\frac{2\pi}{27}}\frac{\alpha_\mathrm{f}}{\beta_T}\frac{\omega_L^2}{\omega}\left[1-\exp{\left(-\frac{T}{\hbar\omega}\right)}\right]g^\mathrm{ff},
\end{equation}
where $u = T/(\hbar\omega)$,  and $g^\mathrm{ff}$ is the Gaunt factor.  In the limit of $\hbar\omega_B\ll kT$ one has
 \begin{equation}
g^\mathrm{ff} = \frac{\pi}{\sqrt{3}}\exp{\left(\frac{T}{2\hbar\omega}\right)}K_0\left(\frac{T}{2\hbar\omega}\right),
\end{equation} 
where $K_0$ is the modified Bessel function, and the expressions for arbitrary magnetic fields could be found in \citep{m40}.

The polarizations of the normal modes can be derived from the dielectric tensor of plasma + vacuum media  $\hat{\epsilon}$, which is related to the matrix $\hat{\Pi}$ \citep{m29}:
\begin{equation}
\hat{\epsilon} = \hat{\mathbf{1}} + \hat{\chi}^\mathrm{(vac)} + \hat{\chi}^\mathrm{(pl)} ; \\
\end{equation}
\begin{equation}
\hat{\chi}^\mathrm{(pl)} = -\frac{\omega_\mathrm{L}^2}{\omega^2}\int\hat{\Pi}f(\beta)\mathrm{d}\beta; 
\label{eq6x}
\end{equation}
\begin{align}
\hat{\chi}^\mathrm{(vac)} &= -2a(\mathbf{k}\otimes\mathbf{k})+4a(\mathbf{k} \mathbf{\times} \mathbf{B})\otimes(\mathbf{k} \mathbf{\times} \mathbf{B}) + 7a (\mathbf{B}\otimes \mathbf{B}),
\label{eq777}
\end{align}
where  $a = \alpha_\mathrm{f}b^2/(45\pi)$. The expression (\ref{eq777}) for permittivity tensor of vacuum is valid for $B \ll B_\mathrm{cr}$. Integrating Eq. (\ref{eq6x}) with the Maxwellian distribution function 
(\ref{eqx4}), we obtain the resulting expression for the  permittivity tensor of the magnetized plasma \footnote{MATLAB code, which evaluates the components of the permittivity tensor and refraction indices for a mildly-relativistic plasma, can be found at http://www.mathworks.com/matlabcentral/fileexchange/52024}
\begin{align}
    \chi_{xx}^\mathrm{(pl)}&=\chi_{yy}^\mathrm{(pl)}=\mathrm{i}(b_{-}+b_{+}), \nonumber \\
    \chi_{xy}^\mathrm{(pl)}&=-\chi_{yx}^\mathrm{(pl)} = b_{-}-b_{+}, \nonumber \\
    \chi_{xz}^\mathrm{(pl)}&=\chi_{zx}^\mathrm{(pl)}=-\mathrm{i}\chi_{yz}^\mathrm{(pl)}=\mathrm{i}\chi_{zy}^\mathrm{(pl)} = c_{-}, \\
    \chi_{zz}^\mathrm{(pl)}&=-\frac{\omega_\mathrm{L}^2}{\omega(\omega+\mathrm{i}\gamma)}+ d_{-}, \nonumber
		\label{eq4}
\end{align}
where 
\begin{align}
    b_{+} &=\frac{\mathrm{i}\omega_\mathrm{L}^2}{2\omega(\omega+\mathrm{i}\gamma +\omega_B)}, \\
    b_{-} &=\frac{\sqrt{\pi}\omega_\mathrm{L}^2}{2\omega^2\beta_T^2}\frac{\varpi(\xi_1)-\varpi(\xi_2)}{\xi_2 - \xi_1}, \\
    c_{-} &= \mathrm{i}\sqrt{\frac{\pi}{2}}\frac{\omega_\mathrm{L}^2\sin\theta}{\omega\omega_B\beta_T}\frac{\xi_1\varpi(\xi_1)-\xi_2\varpi(\xi_2)}{\xi_2-\xi_1}, \\
		 d_{-}&=\frac{\omega_\mathrm{L}^2\sin^2\theta}{\omega_B^2}\left(\frac{\xi_1(1+\mathrm{i}\sqrt{\pi}\xi_1\varpi(\xi_1))-\xi_2(1+\mathrm{i}\sqrt{\pi}\xi_2\varpi(\xi_2))}{\xi_2-\xi_1} + i\sqrt{\frac{\pi}{16}}\frac{b}{\beta_T^2}\frac{\varpi(\xi_1)-\varpi(\xi_2)}{\xi_2 - \xi_1}\right),
\end{align}
\begin{equation}
\xi_{1,2} = \frac{1}{\sqrt{2}\beta_T}\left[\cos\theta\pm\sqrt{\cos^2\theta-2\left(1-\frac{\omega_B}{\omega}+\frac{\hbar\omega}{2m_ec^2}\sin^2\theta+\mathrm{i}\frac{\gamma}{\omega}\right)}\,\right].
\end{equation}

The function
\begin{equation}
\varpi(Z)=\frac{\mathrm{i}}{\pi}\int\limits_{-\infty}^{+\infty}\frac{e^{-y^2}}{Z-y}\mathrm{d}y
\end{equation}
		in the region $\mathrm{Im}(Z)\geq 0$ equals to the Kramp function (the plasma dispersion function), whereas for $\mathrm{Im}(Z)<0$ one can use the relation $\varpi(Z) = -\varpi^*(Z^*)$.
		This function can be easily evaluated numerically with high precision \citep{m24, m25, m26}. Because of the two groups of electrons, which are in resonance with a given photon, the resonant terms in the dielectric permittivity tensor in the mildly-relativistic approximation contain two plasma-dispersion functions, instead of one in the commonly-used tensor for the warm non-relativistic plasma. 
		
Introducing a polarization parameter,
\begin{equation}
q = \frac{-\epsilon_{xx}\cos^2\theta+\epsilon_{yy}+\epsilon_{xz}\sin(2\theta)-\epsilon_{zz}\sin^2\theta}{2i(\epsilon_{xy}\cos\theta+\epsilon_{yz}\sin\theta)},
\end{equation}
and the polarization coefficients of modes,
\begin{equation}
K_\mathrm{X, O} = q\mp\sqrt{q^2+1},
\end{equation}
one can find the polarization of electric field in the X- and O-modes \citep{m29}:
\begin{align}
\label{polar_eqs}
e_x^{(X, O)}&=\frac{K_\mathrm{X, O}\cos\theta}{\sqrt{1+|K_\mathrm{X, O}|^2}}, \nonumber \\ 
e_y^{(X, O)}&=-\frac{\mathrm{i}}{\sqrt{1+|K_\mathrm{X, O}|^2}}, \\ 
e_z^{(X, O)}&=-\frac{K_\mathrm{X, O}\sin\theta}{\sqrt{1+|K_\mathrm{X, O}|^2}}.\nonumber
\end{align}
Here the minus sign in the value $K$ corresponds to the extraordinary (X) mode and plus -- to the ordinary (O) one. {These expressions for the polarization vectors are given in the coordinate system where the x-axis is associated with the transverse component of the wave vector. This is convenient in the case when the magnetic field is perpendicular to the star's surface. In the general case of the inclined magnetic field, it is more convenient to use another reference frame (let us call it x*y*z*), where z*-axis is parallel to the magnetic field and x*-axis lies in the plane that contains both the magnetic field and the normal vector to the surface. In that case a wave vector is rotated by some angle $\phi$ from the x*z*-plane. By} expressing the wave vector $\mathbf{k}$ through the polar angles $\theta$ and $\phi$, we obtain the following expressions for the mode polarization components:
\begin{align}
\label{polar_eqs2}
e_{x*}^{(X, O)}&={e}_{x}^{(X, O)}\cos\phi+{e}_{y}^{(X, O)}\sin\phi, \nonumber \\ 
e_{y*}^{(X, O)}&=-{e}_{x}^{(X, O)}\sin\phi+{e}_{y}^{(X, O)}\cos\phi, \\ 
e_{z*}^{(X, O)}&={e}_{z}^{(X, O)}.\nonumber
\end{align}

We would like to emphasize the importance of xz-, yz-components of the dielectric tensor which are sometimes ignored. These components are responsible for the resonance in the ordinary mode. Their omission leads to incorrect opacity for the ordinary mode in the cyclotron line core for all propagation angles.

\begin{figure*}
    \centering
        \includegraphics[width=1.0\columnwidth]{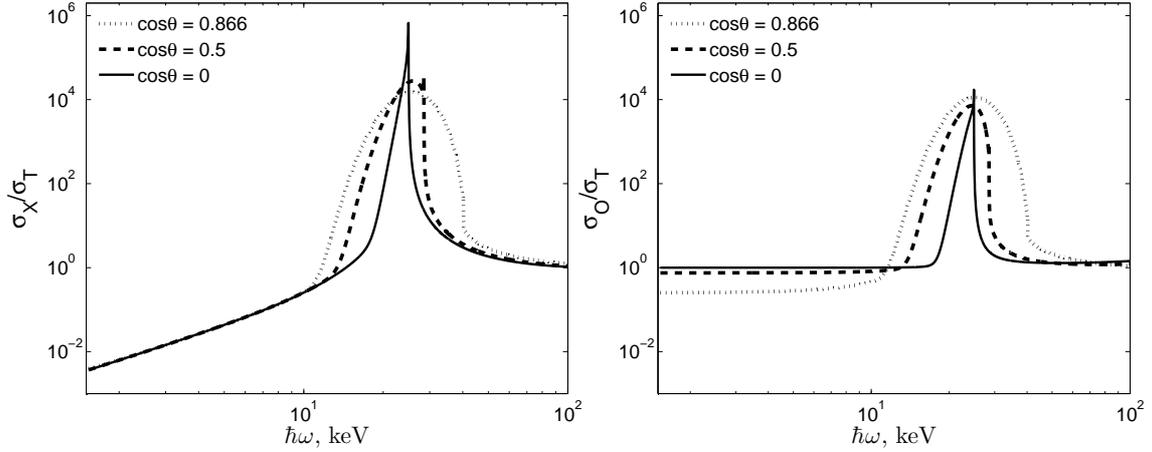}
\caption{Scattering cross sections separately for X-mode (left panel) and O-mode (right panel) for $B = 2.2\cdot10^{12}$\,G, $N_e = 10^{20}$\,cm$^{-3}$ and $T = 20$\,keV.}
\label{fig_dop2}
\end{figure*}

\begin{figure*}
    \centering
        \includegraphics[width=1.0\columnwidth]{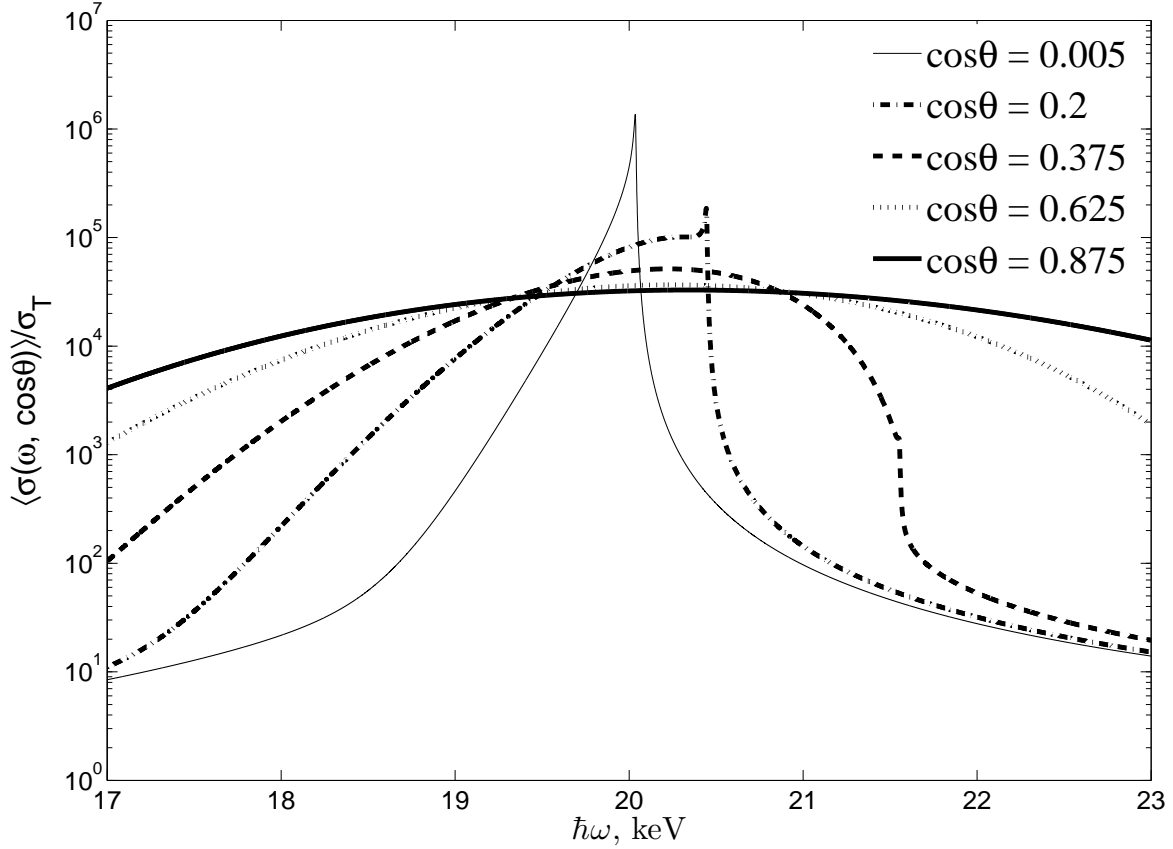}
\caption{Close-up view of the average total scattering cross section near the cyclotron frequency for $b = 0.04$ and $T = 5$\,keV. The narrowest curve corresponds to the almost transverse propagation angle $\cos\theta = 0.005$. Note the sharp cutoff of the resonance curves due to relativistic effects in the case of small values of $\cos\theta$.}
\label{ris1}
\end{figure*}

To compare our results with the results obtained for fully-relativistic case (see, e.g., \citet{m22_y}), we calculate {the scattering cross-sections in the frequency range close to the frequency of the first cyclotron harmonic. It is calculated using Eq. (\ref{eq_sigma}) for $T = 5-20$\,keV and plotted in Fig.~\ref{fig_dop2}-\ref{ris1}}. The shape of the resonance for close to transverse propagation angles is distorted and acquires a narrow feature with the width of the order of $\sqrt{\gamma}$ due to the presence of a cutoff frequency in the resonance condition (\ref{eq3}). Near this point the amount of resonant particles per unit frequency interval tends to infinity if we neglect the line broadening. The asymmetry in the scattering cross-section is especially pronounced  for small values of $\cos\theta$, showing sharper cutoff above the resonance than below it. Note that even at temperatures $T \ll m_ec^2$, the relativistic effects change significantly the scattering cross-section of X-, O-modes propagating nearly transverse to the magnetic field lines at angles $|\theta - \pi/2|\lesssim\beta_T$. 

\section{Transfer equation for the Stokes parameters}

The two equations for the incoherent intensities of the X-, O-modes provide an adequate description of the radiation transfer only if the polarizations of propagating waves evolve adiabatically, i.e., in the absence of the linear mode coupling \citep{m32}. Generally speaking, the geometrical optics (adiabatic) conditions can break at some point on the way out of the atmosphere, and the X-, O-modes become coherent due to the linear mode coupling. In such a case, the transfer equations for modes should be replaced by a complete set of the four transfer equations for the Stokes parameters.  

In what follows, we derive the coefficients in this set of equations in the case of the mildly-relativistic highly magnetized plasma. Although it is possible to obtain the equations using the matrix $\hat{\Pi}$, here we use another technique. For the sake of simplicity, we closely follow the approach of \citet{m11}, where the transfer equations were derived for the cold plasma and without taking into account the frequency redistribution of photons.

In the adopted approach, the propagating electromagnetic waves are described by a vector ${\bf J}$, which is defined in terms of the four Stokes parameters \citep{m29}:
\begin{equation}
\label{eq23}
{\bf J} = \frac{1}{2}\left(
\begin{array}{c}
    I+Q \\ I-Q \\
    2U \\ 2V
\end{array}
\right).
\end{equation}
An evolution of this vector is governed by the transfer equations
\begin{equation}
\label{eq24}
\frac{\mathrm{d}\mathbf{J}}{\mathrm{d}s} = - \hat{M} \mathbf{J} + \mathbf{S}_\mathrm{em} + \mathbf{S}_\mathrm{sc},
\end{equation}
where $s$ is the coordinate along a ray. The source functions ${\bf S}_\mathrm{em}$ and ${\bf S}_\mathrm{sc}$ are the emissivity and photon yields due to scattering from other directions and rays, respectively. The transfer matrix $\hat{M}$ describes the absorption and the escape of radiation due to the scattering as well as the evolution of polarization.  The eigenvectors of the transfer matrix correspond to the X- and O-modes. 

The components of the transfer matrix $\hat{M}$ are related to the components of the dielectric tensor of a media:
\begin{align}
\hat{M} &= \frac{1+\gamma_\mathrm{ff}/\gamma_\mathrm{r}}{2}\left[ 
\begin{array}{cccc}
	2\mathrm{Im}\,\varsigma_{11} & 0 & \mathrm{Im}\,\varsigma_{12} & -\mathrm{Re}\,\varsigma_{12} \\
	0 & 2\mathrm{Im}\,\varsigma_{22} & -\mathrm{Im}\,\varsigma_{12} & -\mathrm{Re}\,\varsigma_{12} \\
	-2\mathrm{Im}\,\varsigma_{12} & 2\mathrm{Im}\,\varsigma_{12} & \mathrm{Im}\,(\varsigma_{11} + \varsigma_{22})& \mathrm{Re}\,(\varsigma_{11}-\varsigma_{22}) \\
	-2\mathrm{Re}\,\varsigma_{12} & -2\mathrm{Re}\,\varsigma_{12} & \mathrm{Re}\,(\varsigma_{22}-\varsigma_{11}) & \mathrm{Im}\,(\varsigma_{11} + \varsigma_{22})
\end{array}\right],
\end{align}
where
\begin{align}
\varsigma_{11}&=\epsilon_{xx}\cos^2\theta+\epsilon_{zz}\sin^2\theta-\epsilon_{xz}\sin{2\theta}, \nonumber\\
\varsigma_{12}&=-\varsigma_{21} = \epsilon_{xy}\cos\theta+\epsilon_{yz}\sin\theta, \\
\varsigma_{22}&=\epsilon_{yy}. \nonumber
\end{align}
Note the terms $\epsilon_{xz}$ and $\epsilon_{yz}$, which are absent in the cold plasma limit.

Now, let us consider an incident electromagnetic wave with the wave vector $\mathbf{k}'$, frequency $\omega'$ and electric field $\mathbf{E}_\mathrm{in}(\mathbf{r}, t) = \mathbf{E}_0\mathrm{e}^{\mathrm{i}\mathbf{k'r}-\mathrm{i}\omega' t}$, where $\mathbf{E}_0 = (E_{0x}\cos\theta'\cos\phi'-E_{0y}\sin\phi')\mathbf{x}^*_0+(E_{0x}\cos\theta'\sin\phi'+E_{0y}\cos\phi')\mathbf{y}^*_0+E_{0x}\sin\theta'\mathbf{z}^*_0$. Here $E_{0x}$ and $E_{0y}$ are the components of the electric field in the reference frame x'y'z', where the z'--axis is directed along the wave vector $\mathbf{k}'$ and x'--axis lies in the plane, containing both $\mathbf{k}'$ and $\mathbf{B}$. A scattered wave, propagating at angles $\theta$ and $\phi$, is characterized by the wave vector $\mathbf{k}$ and frequency $\omega$.
In the dipole approximation, the electric field of the wave scattered by an electron is
\begin{equation}
\mathbf{E}_\mathrm{sc} = \frac{[\mathbf{k}\mathbf{\times}[\mathbf{k}\mathbf{\times}\mathbf{\ddot{d}}]]}{c^2rk^2} = \left[(E_{x}\cos\theta\cos\phi-E_{y}\sin\phi)\mathbf{x}^*_0+(E_{x}\cos\theta\sin\phi+E_{y}\cos\phi)\mathbf{y}^*_0+E_{x}\sin\theta\mathbf{z}^*_0\right]\mathrm{e}^{\mathrm{i}\mathbf{kr}-\mathrm{i}\omega t},
\end{equation}
where $\mathbf{d}$ is the dipole momentum of an oscillating electron, $r$ the distance between an observation point and an electron. Then, for an electron with an undisturbed velocity $\beta$ along the magnetic field, we can construct the matrix, which relates the electric field of the scattered wave to that of the incident wave:
\begin{equation}
   \left(
     \begin{array}{c}
       E_x \\
       E_y
     \end{array}
   \right)_\mathrm{sc} = -\frac{r_e}{r}\left[\begin{array}{cc}
	a_0 & b_0 \\
	c_0 & d_0
	\end{array}\right]\left(
     \begin{array}{c}
       E_{0x} \\
       E_{0y}
     \end{array}
   \right)_\mathrm{in},
\end{equation}
where
\begin{align}
a_0&=\Lambda(\Delta\phi)\cos\theta(\cos\theta'-\beta)+(1-\mathrm{i}\frac{\gamma}{\omega})\sin\theta\sin\theta', \\
b_0&=\Upsilon(\Delta\phi)\cos\theta(1-\beta\cos\theta'), \\
c_0&=\Upsilon(\Delta\phi)(\beta-\cos\theta'), \\
d_0&=\Lambda(\Delta\phi)(1-\beta\cos\theta'), \\
&&\nonumber \\
\Lambda(\Delta\phi)&=\frac{\omega'(1-\beta\cos\theta')}{\omega^{*2}-\omega_B^{*2}}\left(\omega^*\cos{\Delta\phi}+\mathrm{i}\omega_B^*\sin{\Delta\phi}\right),\\
\Upsilon(\Delta\phi)&=\frac{\omega(1-\beta\cos\theta')}{\omega^{*2}-\omega_B^{*2}}\left(\omega^*\sin{\Delta\phi}-\mathrm{i}\omega_B^*\cos{\Delta\phi}\right),
\end{align}
and $\Delta\phi = \phi-\phi'$.

The scattering source function ${\mathbf S}_\mathrm{sc}$ can be expressed in the following way
\begin{equation}
{\mathbf S}_\mathrm{sc} = r_0^2 N_e\int \mathrm{d}\,\Omega'\int\mathrm{d}\omega'\int \mathrm{d}\,\beta f(\beta)\hat{R}({\bf k}'\rightarrow {\bf k}){\bf J}({\bf k}')\delta(\omega-\omega'+\Delta\omega),
\end{equation}
where the scattering matrix is 
\begin{equation}
\hat{R}({\bf k}'\rightarrow {\bf k}) = \left[
\begin{array}[pos]{cccc}
	|a_0|^2 & |b_0|^2 & \mathrm{Re}\,(a_0^*b_0) & \mathrm{Im}\,(a_0^*b_0) \\
	|c_0|^2 & |d_0|^2 & \mathrm{Re}\,(c_0^*d_0) & \mathrm{Im}\,(c_0^*d_0) \\
	2\mathrm{Re}\,(a_0^*c_0) & 2\mathrm{Re}\,(b_0^*d_0) & \mathrm{Re}\,(a_0^*d_0 + b_0^*c_0) & \mathrm{Im}\,(a_0^*d_0 - b_0^*c_0) \\ -2\mathrm{Im}\,(a_0^*c_0) & -2\mathrm{Im}\,(b_0^*d_0)  & -\mathrm{Im}\,(a_0^*d_0+b_0^*c_0) & -\mathrm{Re}\,(a_0^*d_0-b_0^*c_0)  	
\end{array}
\right].
\end{equation}

The emission source function is determined by the requirement that in the thermodynamic equilibrium, when the radiation is unpolarized and its total intensity is equal to the Planck intensity, i.e.,  $\mathbf{J}_\mathrm{eq} = (B_{\omega, T}/2, B_{\omega, T}/2, 0, 0)$, the absorption term should cancel out with the term $\mathbf{S}_\mathrm{em}$.
Thus, one finds that
\begin{equation}
{\mathbf S}_\mathrm{em} = \frac{\gamma_\mathrm{ff} k_0 B_{\omega, T}}{2\gamma_\mathrm{r}}\left(\begin{array}{c} \mathrm{Im}(\varsigma_{11}) \\  \mathrm{Im}(\varsigma_{22}) \\ 0 \\  -2\mathrm{Re}\varsigma_{12} \end{array}\right).
\end{equation}

The equations, derived in this section, are applicable to the transfer of polarized radiation even under the conditions, where the geometrical optics of the X-  and O-modes is not valid and they are coupled due to propagation in an inhomogeneous plasma. 
{The scattering matrix, which we find here, is the simplest possible, which still takes into account all important physical effects. It is known that direct solution to the Stokes equations sometimes requires uncomfortably high spatial resolution to produce accurate results \citep{m11}.  The main reason for that is the oscillations of V and U Stokes parameters in the vicinity of so-called vacuum resonance point in the presence of linear mode coupling, i.e., where the geometrical optics approximation fails.
A method to overcome this difficulty was suggested in \citet{m32_b}, where transfer equations for normal modes are separately solved in two regions before and after the coupling layer and joined together by introducing the ''jump conditions''. This method implies no scattering in the coupling layer. However, as we will show in the next section, the scattering optical depth of the coupling region (under typical conditions for neutron stars) is greater than or of the order of unity for the photons at the cyclotron resonance, so that one has to solve the transfer equations through the coupling layer.}

\section{Linear coupling of the ordinary and extraordinary modes}

The electromagnetic modes cannot be treated independently if their polarization changes too quickly along a ray.
Usually this happens near the point, where the plasma and polarized vacuum contributions to refraction indices are of the same order.  
{This region of so-called vacuum resonance is defined by the condition $|q|^2 \lesssim 1$. There, the rapid change of polarization vectors (\ref{polar_eqs}) in inhomogenous plasma can cause non-adiabatic transformation of modes, that is, linear mode coupling \citep{m11, m32, m42}.}

In a mildly-relativistic plasma the polarization parameter is
{
\begin{equation}
q = \frac{i(b_{-}+b_{+})\sin^2\theta-3a\sin^2\theta + c_{-}\sin(2\theta)-(d_{-}-\omega_L^2/(\omega(\omega+i\gamma)))\sin^2\theta}{2i\left[\left( b_{-}-b_{+}\right)\cos\theta + ic_{-}\sin\theta\right]}
\label{eq_q}
\end{equation}
and can be simplified in the  vicinity of the cyclotron frequency as
}
\begin{equation}
 q \approx \frac{\sin^2\theta}{2\cos\theta}\frac{\mathrm{i}b_{-}-3a}{\mathrm{i}b_{-}},
\end{equation}
{since the terms $c_{-}, d_{-}$, $b_{+}$, and $\omega_L^2/\omega^2$ are much less than the term $b_{-}$. Here $b_{-}$ is due to plasma contribution and $3a$ is due to vacuum contribution. Thus, the necessary condition for the mode coupling in the vicinity of the cyclotron resonance is $|b_{-}| \sim 3a$. For estimation purposes, we consider the case, when $|\cos\theta|\sim 1$.} For such propagation angles
\begin{equation}
b_{-}\approx\sqrt{\frac{\pi}{8}}\frac{\omega_\mathrm{L}^2}{\omega^2\beta_T}\left|\varpi\left(\frac{\omega-\omega_B+\mathrm{i}\gamma}{\sqrt{2}\omega\beta_T}\right)\right|.
\label{b-}
\end{equation}

{Under certain conditions the coupling may be important in the optically thick region. Then the radiation transfer is affected by the mode coupling and the use of Stokes parameters is necessary.}
Let us introduce the scattering optical depth for an X-mode photon with a frequency $\omega$ in a magnetized plasma,
{
\begin{equation}
\tau(\omega) = \int\limits_{-\infty}^z\sigma_{X} N_e(z') \mathrm{d}z', 
\end{equation}
where z is the vertical coordinate in the atmosphere and the axis is directed from infinity to the surface.
For order of magnitude estimates let us use the simple nonrelativistic expression for scattering cross-section in a pure plasma (without vacuum effects; see, for example, \citet{m29, m30})
\begin{equation}
\sigma_{X} \approx \sqrt{\frac{\pi}{2}}\frac{4\pi e^2(1+\cos^2\theta)}{mc\omega \beta_T \cos\theta}\mathrm{Re}\varpi\left(\frac{\omega-\omega_B+\mathrm{i}\gamma}{\sqrt{2}\omega\beta_T\cos\theta}\right).
\end{equation}
Then, substituting $|\cos\theta|$ by unity, we obtain
}
\begin{equation}
\tau(\omega)\approx \sqrt{\frac{\pi}{2}}H\frac{2\omega_\mathrm{L}^2}{c\beta_T\omega}\mathrm{Re}\varpi\left(\frac{\omega-\omega_B+\mathrm{i}\gamma}{\sqrt{2}\omega\beta_T}\right),
\label{tau_eq}
\end{equation}
where {$H$ is the characteristic atmosphere height for an exponential atmosphere with density $N\sim \exp(z/H)$}. 

As a result, equating the plasma and vacuum contributions, {expressing $\omega_L$ via $\tau$  from Eq.\,(\ref{tau_eq}), and substituting it into Eq.\,(\ref{b-})}, we find the optical depth at which the polarization of the X-mode undergoes transition to the vacuum polarization and, hence, the geometrical optics approximation can be broken:
\begin{equation}
\tau_{*}\approx \frac{12\omega H\alpha_\mathrm{f}}{45\pi c}\left(\frac{B}{B_\mathrm{cr}}\right)^2\frac{|\mathrm{Re}\varpi\left(\frac{\omega-\omega_B+\mathrm{i}\gamma}{\sqrt{2}\omega\beta_T}\right)|}{|\varpi\left(\frac{\omega-\omega_B+\mathrm{i}\gamma}{\sqrt{2}\omega\beta_T}\right)|}.
\label{eqlast}
\end{equation}

The largest value of the function $\tau_{*} (\omega)$ is at the cyclotron resonance. If this value exceeds unity, then region of the linear mode coupling lies below the photosphere. For example, in an isothermal hydrogen atmosphere with the Boltzmann density profile one has
\begin{equation}
\tau_{*}(\omega_B) \approx 0.2\left(\frac{T}{50\mbox{\,eV}}\right)\left(\frac{B}{10^{11}\mbox{\,G}}\right)^3\left(\frac{10^{14}\mbox{\,cm\,}\mbox{s}^{-2}}{g}\right)
\end{equation}
near the resonance (here $g$ is the gravitational acceleration).
Thus, for the cyclotron radiation the coupling layer (with possibly non-adiabatic polarization change) can be located in the optically thick regions of atmosphere even in the magnetic fields as low as $10^{11}$\,G, leading to observable effects. At the same time, the linear mode coupling is likely to have a negligible influence on the transfer of continuum photons. 

\begin{figure*}
    \centering
        \includegraphics[width=1.0\columnwidth]{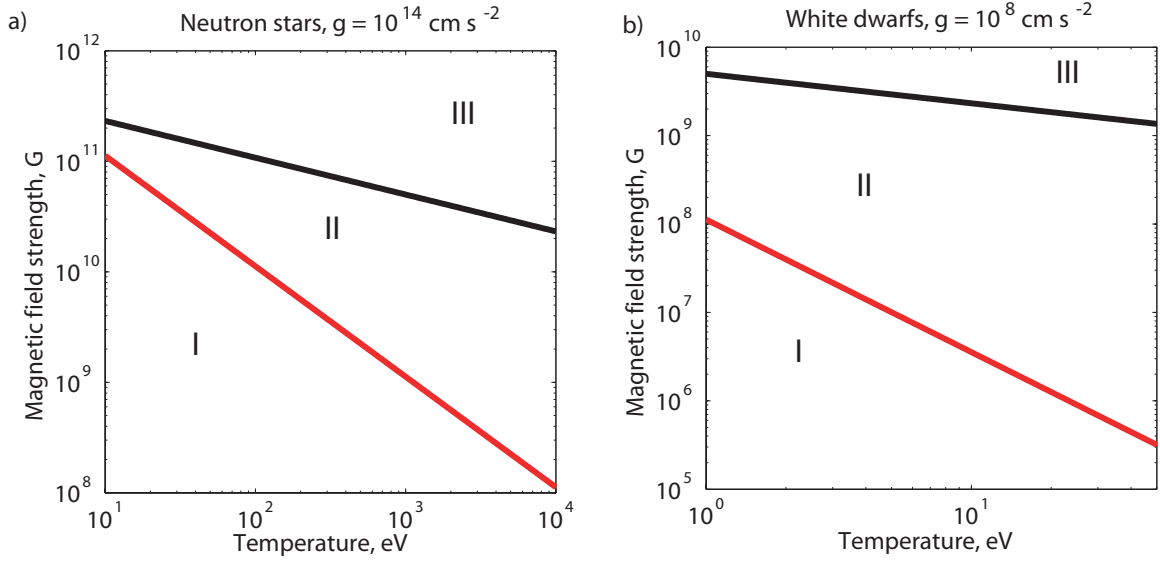}
\caption{The zones of influence for different radiation transfer effects in the parameter space $(B, T)$. Zone I: the photons leave the atmosphere without significant frequency redistribution; in this situation it is sufficient to solve the transfer equations in the quasi-coherent approximation. Zone II: the frequency redistribution plays an important role and determines the cyclotron line profile. Zone III: additionally, the vacuum polarization is strong enough to change the dispersion properties of the X-, O-modes in the regions, where scattering is still important (i.e., below the photosphere, $\tau_{*}(\omega_B)>1$). The panel a: the neutron stars with $g = 10^{14}$ cm\,s$^{-2}$; the panel b: the white dwarfs with $g = 10^8$cm\,s$^{-2}$.}
\label{ris_zones}
\end{figure*}

{Another important phenomenon which affects the radiation transfer in a resonance line is the frequency redistribution during scattering. Regarding cyclotron lines, it was discussed in a number of works \citep{m5, m16, m41}. Here we briefly estimate the atmospheric conditions under which it will play crucial role. Consider a photon, whose probability of escaping from the line due to single excursion into its wing is given by Eq. (7) from \citet{m22}:
\begin{equation}
P_\mathrm{esc} \approx \frac{\beta_T}{\tau\sqrt{8\ln\tau}} \approx \frac{\beta_T}{4\tau}.
\end{equation}
For the frequency redistribution to become important, this probability should be greater than probability of escaping due to spatial diffusion alone,
\begin{equation}
P_\mathrm{esc} > \frac{1}{\tau^2},
\end{equation}
what gives an inequality
\begin{equation}
\tau\gtrsim\frac{\beta_T}{4}.
\end{equation}
This optical depth is to be compared with the so-called thermalization depth \citep{m43}, where most of the escaping photons are generated, 
\begin{equation}
\tau_\mathrm{th} \approx \sqrt{\frac{\gamma_\mathrm{r}}{\gamma_\mathrm{ff}}},
\end{equation}
implying
\begin{equation}
b\gtrsim 5\cdot10^{-8}\sqrt{\frac{10^{14}\mbox{\,cm}\,\mbox{s}^{-2}}{g}}\frac{m_ec^2}{T}.
\label{eq_last}
\end{equation}
  }

Figure~\ref{ris_zones} shows the three zones in the magnetic field -- temperature plane, which call for the different approaches to the radiation transfer problem. Going from lower to higher zone requires taking into account a larger number of effects and solving more complex radiation transfer equations. In zone I, the frequency redistribution of radiation within the cyclotron line core is weak, so that the major fraction of emerging photons leaves the atmosphere due to usual diffusion. The radiation transfer in this situation can be considered in the quasi-coherent approximation. In zone II, the frequency redistribution of radiation in the line's core is strong, and photons primarily leave the atmosphere due to excursions to the line's wings rather than due to the spacial diffusion. This zone is bound by the condition (\ref{eq_last}), which ensures that most of the resonant photons in emergent spectra have at least one excursion to the line's wings during their passage through the atmosphere. In this zone one should use more complicated transfer equations for incoherent X-, O-modes with partial frequency redistribution effects taken into account. Additionally, in the zone III, the vacuum polarization effects become important in the inner regions of atmospheres (i.e., where $\tau_{*}(\omega_B)>1$) and may lead to the linear mode coupling. In this case, the transition between the plasma-determined polarization and the vacuum-determined polarization occurs in the optically thick regions of the atmosphere, forcing one to use the complete set of transfer equations which can deal with non-adiabatic propagation of modes, i.e., the equations for the Stokes parameters. 


\section{Conclusions}

We analyse, for a wide range of temperatures and magnetic field strengths, the impact of the mildly-relativistic and QED effects on the radiative transfer of the gyro-resonant radiation in the atmospheres of compact stars. We
find that the parameter space can be split into three zones with different requirements to the amount of details in the radiation transfer modeling. In the first zone the simplest and well-known quasi-coherent approximation is valid, which reduces by one the effective number of dimensions in the transfer equations. The second zone requires the mildly-relativistic effects to be taken into account as they determine the frequency redistribution of cyclotron radiation out of the line's core. In this case a system of two equations for the ordinary and extraordinary mode intensities can still be used, and we derive the proper coefficients in these equations. The third zone calls for the most comprehensive analysis, with both the mildly-relativistic and QED effects taken into account. In this case, the linear mode coupling necessitates to use the complete set of the four transfer equations for the Stokes parameters. 
{However, even in a general case, the latter equations have to be used only in the region of linear mode coupling, i.e., close to the point of vacuum resonance, and outside this region the equations for incoherent X-, O-modes are valid.}

The complexity of the radiation transfer modeling in the third zone is due to an inhomogeneity of plasma in a region of the 'vacuum resonance', where the modes change their appearance from the plasma-determined polarization to the vacuum-determined polarization. For the neutron star atmospheres and at the frequencies near the cyclotron resonance, the transition region can be located at the optical depths larger than unity, even for the relatively weak magnetic fields on the order of $10^{11}$\,G. The effect of the linear mode coupling can lead to the observable features in the emergent spectra, though its detailed analysis is a subject of future work and will be presented in a follow-up paper.

In the present paper, we derive the coefficients in the transfer equations both for the X-, O-mode intensities and for the Stokes parameters. {The coefficients in this form are necessary and sufficient for solving the radiation transfer problem for the corresponding regions in the plasma parameter space. The derived equations take into account} the effects of the vacuum polarization, inclined magnetic field, and relativistic corrections to the cyclotron resonance condition. A detailed analysis of these effects, both in the isolated neutron stars and accreting X-ray pulsars, is hindered by a high complexity of the problem. However, a progress in observations and a rapid growth of the available computing resources will allow one to use the results, obtained in this paper, for building realistic models of the atmospheres of such objects in the near future.

We also note that all physical effects analysed in this paper can be important for the atmospheres of magnetized white dwarfs too, though the influence of the vacuum polarization and the subsequent linear mode coupling is significant only for the extreme white dwarfs with a magnetic fields exceeding $10^9$\,G, which are not observed yet.

For those readers, who are interested in the numerical analysis of the dispersion properties of the mildly-relativistic magnetized plasma, we publish the MATLAB code, which evaluates the components of the permittivity tensor and refraction indices. The code can be found at http://www.mathworks.com/matlabcentral/fileexchange/52024 .

\section*{Acknowledgments}

This work was supported by the Government of the Russian Federation (Project No. 14.B25.31.0008).





\bibliographystyle{mnras}
\bibliography{mnras_biblio}
\bsp
\label{lastpage}
\end{document}